Article

# Metadynamics sampling in atomic environment space for collecting training data for machine learning potentials

Dongsun Yoo[1,2, §], Jisu Jung[1, §], Wonseok Jeong[1], and Seungwu Han[1,*]

[1]Department of Materials Science and Engineering and Research Institute of Advanced Materials, Seoul National University, Seoul 08826, Korea.

[2]Present address: Samsung Display Co., Ltd., 1, Samsung-ro, Giheung-gu, Yongin-si, Gyeonggi-do 17113, Korea.

§ Dongsun Yoo and Jisu Jung contributed equally to this paper.

*Corresponding author: hansw@snu.ac.kr

# ABSTRACT

The universal mathematical form of machine-learning potentials (MLPs) shifts the core of development of interatomic potentials to collecting proper training data. Ideally, the training set should encompass diverse local atomic environments but the conventional approach is prone to sampling similar configurations repeatedly, mainly due to the Boltzmann statistics. As such, practitioners handpick a large pool of distinct configurations manually, stretching the development period significantly. Herein, we suggest a novel sampling method optimized for gathering diverse yet relevant configurations semi-automatically. This is achieved by applying the metadynamics with the descriptor for the local atomic environment as a collective variable. As a result, the simulation is automatically steered toward unvisited local environment space such that each atom experiences diverse chemical environments without redundancy. We apply the proposed metadynamics sampling to H:Pt(111), GeTe, and Si systems. Throughout the examples, a small number of metadynamics trajectories can provide reference structures necessary for training high-fidelity MLPs. By proposing a semi-automatic sampling method tuned for MLPs, the present work paves the way to wider applications of MLPs to many challenging applications.

# INTRODUCTION

By delivering the accuracy of density-functional theory (DFT) calculations at much lower costs, atomistic simulations based on machine-learning potentials (MLPs) are being established as a new pillar in computational material science.[1] Most MLPs utilize the locality of quantum systems and so the computational cost increases linearly with respect to the system size, which is a significant advantage over DFT with a cubic scaling.[2] Until now, various types of MLPs have been proposed; neural network potential (NNP),[3] Gaussian approximation potential (GAP),[4] moment tensor potential,[5] deep tensor neural network,[6] and gradient-domain machine learning.[7] In particular, the NNP and GAP are garnering wide interests with applications to challenging simulations such as crystallization behaviors of GeTe,[8,9] and $Ge_2Sb_2Te_5$,[10] Ni-silicidation process,[11] proton transfer at the ZnO-water interface,[12] structure search of $Pt_{13}H_x$ clusters,[13] crystal structure prediction,[14] and identification of active sites in bimetallic catalysts for $CO_2$ reduction.[15]

At the heart of the traditional classical potential is the mathematical formula that captures underlying bonding natures. In contrast, universal mathematical structures of MLPs shift the core of potential development to collecting a proper training set that defines atomic environments wherein the trained MLP is valid. Ideally, the training set should encompass diverse local configurations that may appear in target simulations. In usual practices, the training set is selected based on crystal-derived structures and their molecular dynamics (MD) trajectories. However, MD simulations are conditioned by the Boltzmann statistics, which over-represents low-energy regions and can sample only a few distinct configurations separated by low thermal barriers. As a result, a large pool of reference structures is handpicked manually, which demands expertise from practitioners as well as several iterative refinements of MLPs.[16–18] We note that some methods such as random structure search[19] and entropy-maximization[20] focus on sampling diverse configurations, but they have not been employed in complicated simulations as far as we are aware.

The above discussions call for a sampling method specifically aiming to prepare training sets for MLPs, which is tuned to collect local atomic environments as diverse as possible within the time- and size-scale of DFT calculations. In addition, the sampled configurations should be relevant in the intended simulations. We herein propose one such approach based on metadynamics.[21] The metadynamics defies the Boltzmann distribution by accumulating bias potentials along the collective variables (CVs). Instead of usual implementations that formulate CVs from a set of atomic positions in the real space,[22] we employ as CVs the coordinates in the abstract atomic-environment space, which is spanned by the atom-centered symmetry-function vector (**G**).[23] Being widely used as input features of NNPs, the **G** vector parametrizes local atomic environments into fixed-length vectors via integrating radial and angular distributions of neighboring atoms. By accumulating bias potentials in the **G** space, the present metadynamics (abbreviated as G-metaD hereafter) drives each atom to evolve towards unvisited points in the **G** space. In addition, G-metaD is controlled by a few hyperparameters and can start with simple initial structures, requiring less expertise than the conventional MD-based sampling. Recently, J. Herr et al. suggested a metadynamics sampling method using the distance matrix of the whole system as a CV, which enhanced the stability of MD simulations compared to conventional MD sampling.[24] However, individual atoms in this approach can still repeatedly encounter similar local environments because the CV is based on the total system configuration.

In the following section, we formulate the G-metaD and demonstrate its applications to three systems: H:Pt(111), GeTe, and Si. The first model, H:Pt(111), is chosen to directly compare the sampling style between the conventional MD and G-metaD. The other two examples, GeTe and Si, were studied by using the NNP or GAP.[8,9,17,25,26] We choose these materials to benchmark the performance of NNPs trained by G-metaD trajectories against the state-of-the-art MLPs trained over a large number of structures that were prepared manually.

## RESULTS

**Metadynamics simulation**

The present G-metaD employs the **G** vector as the CV. The local bias potential ($u_b$) is defined as a function of **G**, and the summation of atomic local biases constitutes the total bias potential ($U_b$) applied on the system:

$$U_b(\{\mathbf{R}(t)\}) = \sum_{i=1}^{N_{at}} u_b(\mathbf{G}_i(t)), \qquad (1)$$

where $\{\mathbf{R}(t)\}$ and $\mathbf{G}_i(t)$ are the set of position vectors and the symmetry-function vector of the $i$th atom at time $t$, respectively, and $N_{at}$ is the number of atoms in the system. The biasing force on each atom is computed as follows:

$$F_{i,\alpha} = -\frac{\partial U_b}{\partial R_{i,\alpha}} = -\sum_{j=1}^{N_{at}}\sum_{s=1}^{N_G} \frac{\partial U_b}{\partial G_{j,s}} \frac{\partial G_{j,s}}{\partial R_{i,\alpha}}, \qquad (2)$$

where $F_{i,\alpha}$ and $R_{i,\alpha}$ are the $\alpha$-component ($\alpha = x$, $y$, and $z$) of the force and position vectors of the $i$th atom, respectively, and $G_{j,s}$ is the $s$th component of $\mathbf{G}_j$ with the dimension of $N_G$. For multi-component systems, $u_b$ is defined independently for each atomic species.

We construct the local bias $u_b$ in Eq. (1) using Gaussians centered at **G** points visited by each atom. Since the elements of **G** vectors are highly correlated to each other, it is ineffective to adopt isotropic Gaussians with a fixed width. This is illustrated in Fig. 1a, which schematically shows a typical distribution of training points (grey dots) along two components ($G_i$ and $G_j$). To sample the distribution with isotropic Gaussian biases (black dots with circles whose radius means the Gaussian width), many Gaussians should be accumulated because of the highly anisotropic distribution. To overcome this problem, we employ geometry-adapted Gaussians, which was developed to reconstruct the accurate potential energy surface (PES) from metaD by adjusting the shape and size of Gaussian biases according to the distribution of visited CVs.[27] The present G-metaD reformulates this approach and accumulates adaptive Gaussians centered at visited **G** points as follows:

$$u_{\mathrm{b}}(\mathbf{G}(t)) = h \int_0^t \sum_{i=1}^{N_{\mathrm{at}}} \sum_{n \in \mathbb{N}} \exp\left[-\frac{1}{2}(\mathbf{G}(t) - \mathbf{G}_i(t'))^{\mathrm{T}} \Sigma(\mathbf{G}_i(t'))^{-1}(\mathbf{G}(t) - \mathbf{G}_i(t'))\right] \delta(t' - n\tau) dt'. \quad (3)$$

The covariance matrix Σ in Eq. (3) is given by:

$$\Sigma_{jk}(\mathbf{G}) = \sigma^2 \sum_{i=1}^{N_{\mathrm{at}}} \sum_{\alpha=x,y,z} \frac{\partial G_j}{\partial R_{i,\alpha}} \frac{\partial G_k}{\partial R_{i,\alpha}} + \varepsilon \delta_{jk}, \quad (4)$$

where $G_j$ and $G_k$ are the *j*th and *k*th components of **G**, respectively. In Eqs. (3) and (4), hyperparameters *h*, *σ* and *τ* represent the height and width of Gaussian potentials and time interval of bias updates, respectively. The high correlations among components of **G** render $\Sigma^{-1}$ to be numerically unstable. To prevent the divergence, a small regularization term *ε* (fixed to $10^{-4}$) is added to the diagonal components in Eq. (4). According to Eqs. (3) and (4), the width of the Gaussian bias is adjusted anisotropically such that the Gaussian shape resembles the data distribution as shown in Fig. 1b. This enables the G-metaD to search the relevant regions with much smaller number of bias potentials than in Fig. 1a.

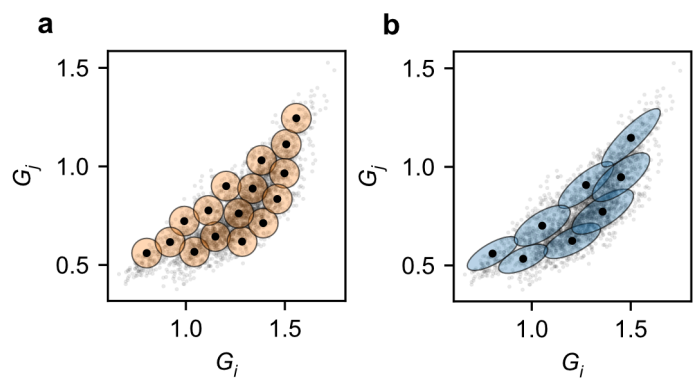

**Fig. 1** The schematic description of bias potential in the **G** space projected on certain components $G_i$ and $G_j$. **a** Isotropic Gaussian potentials with a fixed width. **b** Adaptive Gaussian potentials. Grey dots represent the **G** points sampled during the MD simulation. Black dots indicate the **G** points where bias potential is evaluated and semi-transparent circles represent the bias potentials centered on black dots.

There are three hyperparameters in G-metaD: $h$, $\sigma$, and $\tau$. Being related to the bias strength, $h$ and $\sigma$ control the height and width of Gaussian potentials, respectively. In the case of $h$, a proper value is chosen such that the magnitude of $U_b$ has an order similar to the thermal energy and the simulation remains to be stable. For $\sigma$, if it is too small, a large number of bias potentials would be needed to fill a basin of the PES. In contrast, a large $\sigma$ can obscure the curvature of the PES, thereby causing under-sampling. We find that $\sigma$ around 1 Å is a reasonable choice. Lastly, $\tau$ should be long enough for the system to respond to the updated bias force, and short enough that the metadynamics trajectories can search diverse configurations in the limited simulation time. Our experiences indicate that 20 fs is a sound choice for $\tau$, which is used throughout the present work. Note that these hyperparameters could be assigned differently for each type of atom.

We implement a pair style that computes the bias potential into the LAMMPS package[28] using the SIMPLE-NN library.[29] By operating the client-server mode, one can interface LAMMPS with other *ab initio* codes such as VASP[30] to perform the G-metaD.

**Hydrogen on the Pt(111) surface**

To compare G-metaD and conventional MD in terms of the sampling style, we investigate diffusion of the H atom on the Pt(111) surface. Noble metals are widely used as efficient catalysts for H-involved reactions such as hydrogen evolution reactions[31] and $CO_2$ reduction.[32] An accurate description of the H diffusion on the metal surface is important for simulating these reactions. Furthermore, it has been reported that H atoms can diffuse into the subsurface, influencing the total diffusion kinetics and reaction rates.[33,34] Thus, sampling various H sites on the surface as well as in the subsurface would be important for training MLPs that aim to simulate catalytic reactions.

To obtain the training data, we carry out three simulations on (3×3)-Pt(111) with one H atom adsorbed on the surface; standard MD at 600 and 1700 K and G-metaD at 600 K under NVT conditions. In the case of G-metaD, the bias potential is applied only on the H atom with $h$ and $\sigma$ of 24 meV and 1.0 Å, respectively. The total simulation time is 3 ps for every simulation. The detailed set up for DFT calculations are presented in the Methods section and simulation movies are provided as Supplementary Videos 1-3. At lower temperatures than 1700 K, say, 1500 K, the diffusion into subsurface is not observed during 3-ps MD simulations. In Fig. 2a, the trajectories of the H atom are classified into four regions; face-centered cubic (fcc), hexagonal (hex), bridge and top sites. At 0 K, the lowest energies within each region are 0 (fcc; the reference site), 59 (hex), 47 (bridge), and 37 (top) meV. The lowest energy site in the sublayer is the tetrahedral site right below the top site with the energy of 823 meV in reference to the fcc site.

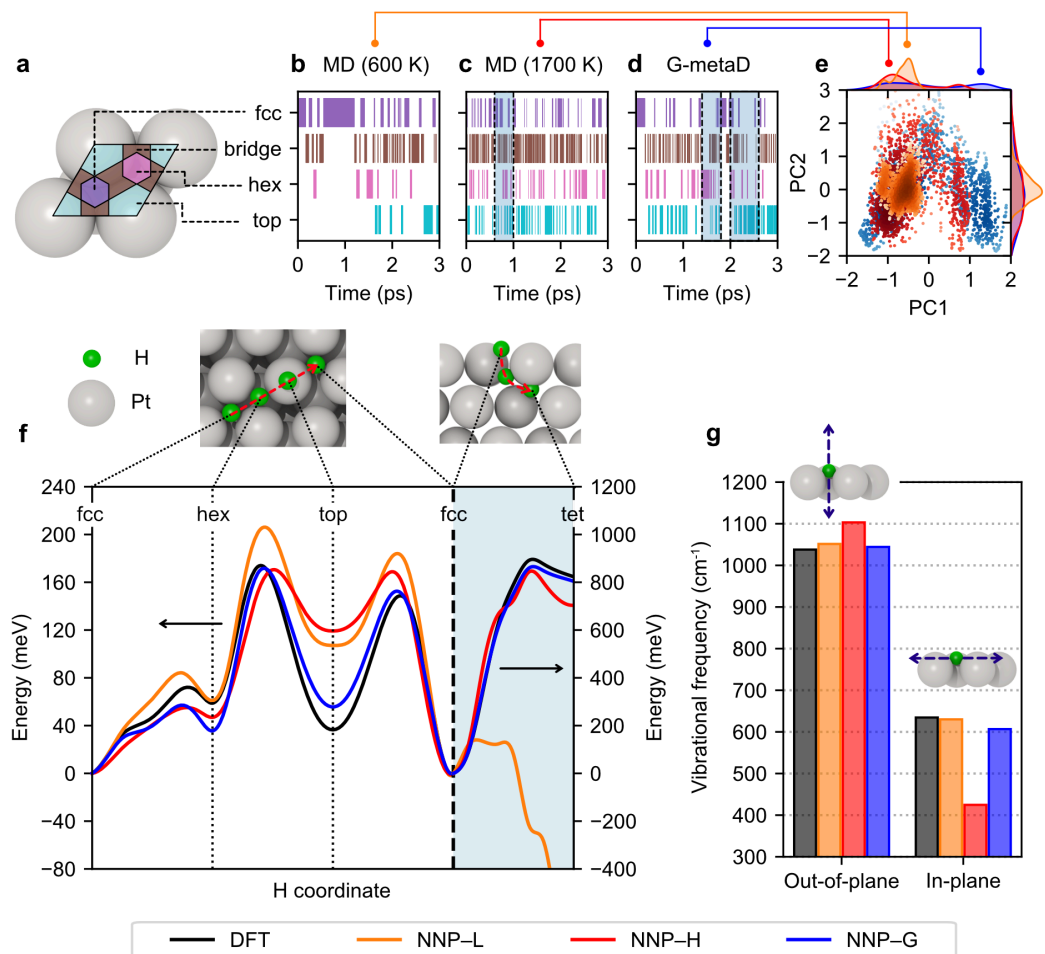

**Fig. 2** Sampled configurations and performances of the corresponding NNPs. **a** Characteristic area on Pt(111) surface. **b-d** The classification of H sites along the time for MD at 600 K, 1700 K, and G-metaD at 600 K, respectively. The shaded area means that the H atom is on the sublayer. **e** Distribution of sampled points of H atom on principal-components axes. **f** The minimum energy paths along fcc → hex → top → fcc → tet (subsurface tetrahedral site). **g** Vibrational frequencies of the H atom at the fcc site. (The atomic configurations are visualized by OVITO.[39])

The temporal evolution of visited sites are displayed in Figs. 2b–d. At 600 K, the H atom stays mostly at the fcc site that is the lowest in the potential energy, which is consistent with the Boltzmann distribution. In addition, the H atom does not penetrate into the subsurface due to a high diffusion barrier of 0.9 eV (see below). At the elevated temperature of 1700 K in Fig. 2c, various sites are sampled more or less evenly and subsurface diffusion is observed (shaded area). On the other hand, G-metaD at 600 K also samples various sites including the sublayer (see Fig. 2d). The H atom in the G-metaD stays within the sublayer for ~1 ps out of 3-ps simulation time, in contrast to ~0.5 ps duration in 1700-K MD (see Fig. 2c). Figure 2e shows the distribution of the three trajectories projected onto major principal axes from principal-component analysis. It is seen that the G-metaD covers a wider area than 600- or 1700-K MD.

Next, we train three NNPs by employing trajectories from each simulation as training data. (The trajectories are sampled every 10 fs.) They are named NNP-L, NNP-H, and NNP-G according to the MD types (600-K MD, 1700-K MD, and G-metaD, respectively). The training errors are similar among the three NNPs and root-mean-squared errors (RMSEs) for energy and force are less than 3 meV/atom and 0.2 eV/Å, respectively. (We refer to the Methods section for the details of NNPs and training procedures.) To compare accuracy of the trained NNPs, we compute in Fig. 2f the minimum energy paths (MEPs) between the three symmetric sites (fcc, hex, and top) using the nudged-elastic-band method using 9 replicas between the symmetric sites.[35,36] The MEP into the subsurface tetrahedral site is also calculated on the right side. For reference, DFT results are also presented, which agree well with literature.[37,38] Overall, the results obtained with the NNP-G best agree with the DFT results except that NNP-G incorrectly estimates the energy of the hexagonal site to be more stable than that of the top site (36 vs. 55 meV). It is notable that both NNP-L and NNP-H show large errors of ~0.1 eV at the top site. The undersampling in 600-K MD for this site (see Fig. 2b) would be responsible for the error with NNP-L. Even though top sites are well sampled in MD at 1700 K (see Fig. 2c), the large error implies that the trajectory fails to capture the low-energy surface because the MD is heavily influenced by wide atomic vibrations. The same reasons account for the errors along the diffusion into the subsurface (shaded area of Fig. 2f); a dramatic failure of NNP-L certainly originates from the absence of data in this region (see Fig. 1b), which is partly resolved by MD at 1700 K. However, a substantial error of ~0.1 eV remains at the subsurface tetrahedral site. This implies that the high-temperature sampling, while useful for overcoming energy barriers, risks undersampling atomic environments around local minima due to the wide vibrations and entropic effects. In contrast, the G-metaD is performed at moderate temperatures so it does not suffer from such problems.

Figure 2g compares vibrational frequencies at the stable fcc sites. There are two independent vibrational modes, out-of-plane and in-plane (twofold). The accuracy with respect to DFT results follows the order of NNP-L > NNP-G > NNP-H. This is consistent with the above observations: the 600-K MD most densely samples the fcc site, resulting in the highest accuracy. In contrast, 1700-K MD undersamples this site because of large thermal energies. The reasonable accuracy with the NNP-G implies that the basin of the PES was sampled sufficiently.

**Amorphization of GeTe**

The present G-metaD is controlled by two hyperparameters, $h$ and $\sigma$. By tuning these two parameters, one can steer the system to explore different regions in the **G**-space, which enables a semi-automatic sampling. We demonstrate this with GeTe, an archetypal phase-change material that has been extensively studied for non-volatile memory devices.[40] Several studies employed NNPs in simulating the amorphous structures and crystallization behaviors of GeTe.[8,25,26] To sample diverse local orders, the training set included liquids, crystals, amorphous phases as well as quenching trajectories. To improve the stability of the simulation, non-stoichiometric phases were also considered.[2,8,25] Here we attempt to prepare the training set for simulating GeTe by utilizing G-metaD only.

Starting from the crystalline rock-salt structure, the G-metaD for GeTe is carried out for 20 ps under NPT conditions of 64 atoms, 0 kbar, and 600 K. For simulating the whole melt-quench process as well as crystallization behaviors, it is necessary to sample both liquid structures with high energies and amorphous structures with the local order similar to those in the crystalline phases. To this end, we generate four G-metaD trajectories with different choices of ($h$, $\sigma$): (8.0, 1.5), (8.0, 1.0), (0.8, 1.0), and (0.8, 0.5) in (meV, Å). In Fig. 3a, the evolution of potential energies during 20-ps G-metaD is shown for each ($h$, $\sigma$). (The movies for G-metaD trajectories are provided as Supplementary Videos 4-7.) To sample local order close to the crystalline structure, we restart G-metaD at 10 ps from the rock-salt structure while maintaining the bias potential accumulated during the preceding 10 ps to avoid redundancy. Here we use $h$ values of 8 or 0.8 meV, which are far smaller than 24 meV in the previous example. This is because there are 32 atoms that contribute to the bias potential (see Eq. (3)) while there was only one atom (H) in the previous case. It is seen in Fig. 3a that at the strongest bias ($h$ = 8 meV and $\sigma$ = 1.5 Å), the system widely changes and even phase separations are noticeable near the end of the G-metaD (see inset figures at the top). In Ref. 8, diffusional mixing of liquid Ge and Te was considered to prevent unphysical phase separations from the *ad hoc* energy mapping. Such atomic environments are automatically sampled in the strongly biased G-metaD. In contrast, under the weakest bias strengths ($h$ = 0.8 meV and $\sigma$ = 0.5 Å), the trajectory remains relatively close to the crystalline structures and appears to mainly sample amorphous-like structures.

To analyze characteristic structures that each G-metaD samples, we introduce the Mahalanobis distance, which is used in measuring distances between a point and distributions.[41,42] (See the Methods section for details.) Using the Mahalanobis distance, we classify atomic environments into crystal, amorphous, and liquid structures. If the distance does not satisfy the given criteria for any of the three phases, the sampled **G** point remains to be unclassified. Histograms in Fig. 3b display the phase fractions sampled for each ($h$, $\sigma$). At the strongest bias, unclassified structures are the most dominant. The rapidly accumulating bias potentials drive the system to evolve towards high-energy structures resembling surfaces or unmixed phases. As the bias strength is reduced, the relative portions of bulk structures, in particular rock-salt structures, increases. This analysis indicates that the system can explore distinct regions in the **G** space by tuning the hyperparameters.

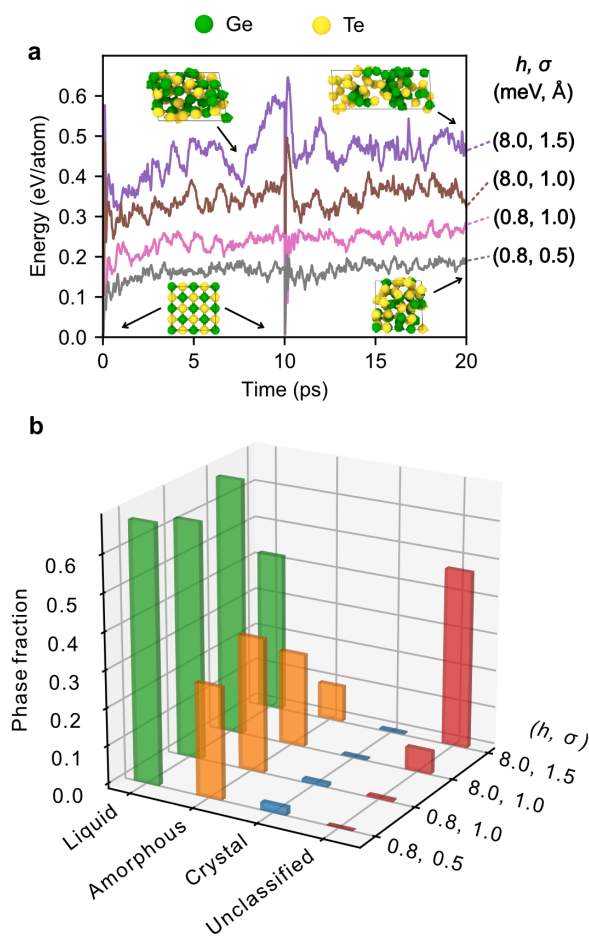

**Fig. 3** Energy and fraction of phases in the four G-metaD. **a** Time evolution of potential energy of four

G-metaD along the time. The energy is referenced to that of the rock-salt structure. Structures during G-metaD are shown as inset figures. **b** The phase fractions classified by Mahalanobis distances shown for each ($h$, $\sigma$).

Using the four G-metaD trajectories, we train an NNP with the energy, force, and stress RMSEs of 6 meV/atom, 0.24 eV/Å, and 5 kbar, respectively. (Trajectories are sampled every 20 fs.) In Fig. 4a, the equation of states (EOS) for rock-salt (*Fm3m*) and rhombohedral (*R3m*) phases are compared between NNP and DFT. The equilibrium volume and bulk moduli agree with DFT within 1%. Even though the EOS and deformed crystals were not explicitly included in the training set, good agreements are found with both phases. This indicates that G-metaD can automatically sample various lattice distortions around the equilibrium structure. However, the small energy difference between rock-salt and rhombohedral phases (8 meV/atom) is neglected by the NNP (< 1 meV/atom).

Next, we perform melt-quench simulations with the NNP and characterize structural properties of resulting liquid and amorphous structures. To compare with DFT on an equal footing, we select a 96-atom supercell for the simulation. The temperature protocol is identical to that in Ref. 8. During the melt-quench process, we do not observe any artefacts such as phase separations. Figures 4b and 4c compare the total and element-resolved radial-distribution functions (RDFs) for liquid and amorphous phases. Overall, good agreements with DFT are found, which is comparable to the previous studies.[8] In Fig. 4d, we analyze the ring statistics by using the R.I.N.G.S. code.[43] Albeit overestimated in densities, the overall ring distributions including the portion of ABAB-type four-membered rings are similar between NNP and DFT. We also simulate the crystallization behavior of a 4096-atom supercell at 500 K with an atomic density of the amorphous phase. (See Fig. 5e.) The crystallization speed is similar to those in Refs. 8 and 9. In Ref. 8, it was found that NNP tends to produce flat four-fold rings, resulting in unphysically fast crystallization, which was improved when relaxation paths from the flat to puckered four-fold rings were included in the training set. The present NNP produces flatness in between the conventional and refined NNPs (not shown), implying that the fine details of medium-range order are not well captured by either MD or G-metaD.

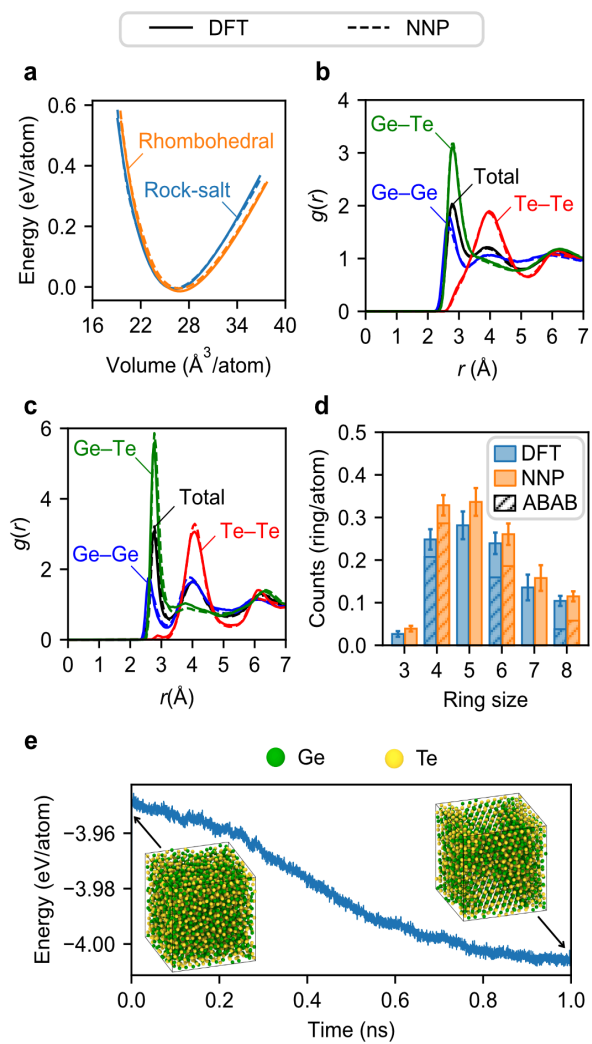

**Fig. 4** Structural properties and crystallization behavior of GeTe obtained by the NNP. **a** The energy-volume relation for rock-salt and rhombohedral phases. The energy is referenced to that of the rock-salt phase at the equilibrium. **b-c** The total and element-resolved radial distribution functions ($g(r)$) for liquid at 1000 K (b) and amorphous structures at 300 K (c). In **a-c**, solid and dashed lines indicate DFT

and NNP results, respectively. **d** Ring statistics of amorphous GeTe at 300 K. The error bars indicate one standard deviation obtained from four independent MD simulations. The ABAB-type rings (A = Ge and B = Te) for even-membered rings are also shown. **e** The time evolution of potential energy during the crystallization. Initial and final structures are shown as inset figures.

**General-purpose potential for Si**

Due to the limited transferability, most MLPs are trained for specific applications. Developing general-purpose MLPs is a formidable task involving construction of a huge data set that contains a vast range of chemical environments, which in turn requires deep understanding of the system and possibly several iterations of refinements of MLPs. There have been a few attempts to generate general-purpose MLPs with manually selected data set.[17,18] For example, in Ref. 17, a general-purpose GAP potential was developed for Si with the training set covering a long list of distinct structures such as several polymorphs, extended and point defects, slab models, amorphous and liquid phases (29 in total).

Here we demonstrate with Si that the G-metaD can be applied to preparing a training set for the general-purpose potential in convenient ways. Using two sets of hyperparameters ([$h$ (meV), $\sigma$ (Å)] = [0.4, 1.0] and [0.04, 1.0]), we generate two 30-ps G-metaD trajectories starting from a 64-atom supercell in the cubic-diamond (cd) structure, under the NPT condition (0 kbar and 600 K). Like in GeTe, G-metaD restarts from the crystalline Si every 10 ps. Here we use smaller $h$ values than the previous GeTe example because of the larger number of atoms contributing to the bias (64 vs. 32) and smaller degrees of configurational freedom (unary vs. binary). The simulation time is extended to 30 ps to sufficiently sample local environments far from the diamond structure such as high-temperature liquids and under-coordinated atoms. (The movies for G-metaD trajectories are provided as Supplementary Videos 8 and 9.) Using the two G-metaD trajectories sampled every 20 fs, we train an NNP with the energy, force, and stress RMSEs of 23 meV/atom, 0.36 eV/Å, and 5 kbar, respectively.

Figure 5a enlists elastic, surface, and defect properties of Si computed with the NNP, which are scaled by the DFT values. These properties were selected by Ref. 17 in benchmarking the general-purpose potential of Si. Overall, NNP results show reasonable agreements with DFT although the test structures were not explicitly included in the training set. The mean absolute error for these properties is 11%, higher than 6% in Ref. 17 where the data set was constructed manually. To note, planar defects were not included in the training data of Ref. 17, and errors of the unstable stacking fault energies for shuffle ($\gamma_{us}^{(s)}$) and glide ($\gamma_{us}^{(g)}$) are −16% and 13%, respectively, which are larger than −11% and 5% in the present work.

Figure 5b shows EOS of various phases of Si. We note significant deviations for high-pressure phases such as hcp and bc8. During G-metaD simulations, atoms in the same supercell are driven to different chemical environments (i.e., **G** vectors), unlike crystalline structures where atoms share a few **G** vectors. Therefore, the prediction error tends to increase in crystalline structures with local orders different from the initial one. This can be improved by augmenting the training set (see the Discussion section). On the other hand, the structural properties of liquid and amorphous phases in Figs. 5c–f are in reasonable agreement with DFT except that the tetrahedral unit in the amorphous Si (*a*-Si) is more rigid with the NNP (see Fig. 5f).

Lastly, we calculate in Fig. 5g energies of Si nanoclusters with the number of atoms ranging over 4–20.[44–46] We adopt geometries from Refs. 44, 45 and relax them using DFT and NNP. (The structures in the inset are obtained by NNP.) It is seen that the NNP describes energies of the small-size clusters well except for the smallest 4-atom cluster. This indicates that the uncoordinated Si atoms are also sampled within the G-metaD trajectories with the high bias. However, the present NNP may not provide enough accuracy to delineate the energy ordering among various geometries of the same-number nanoclusters.

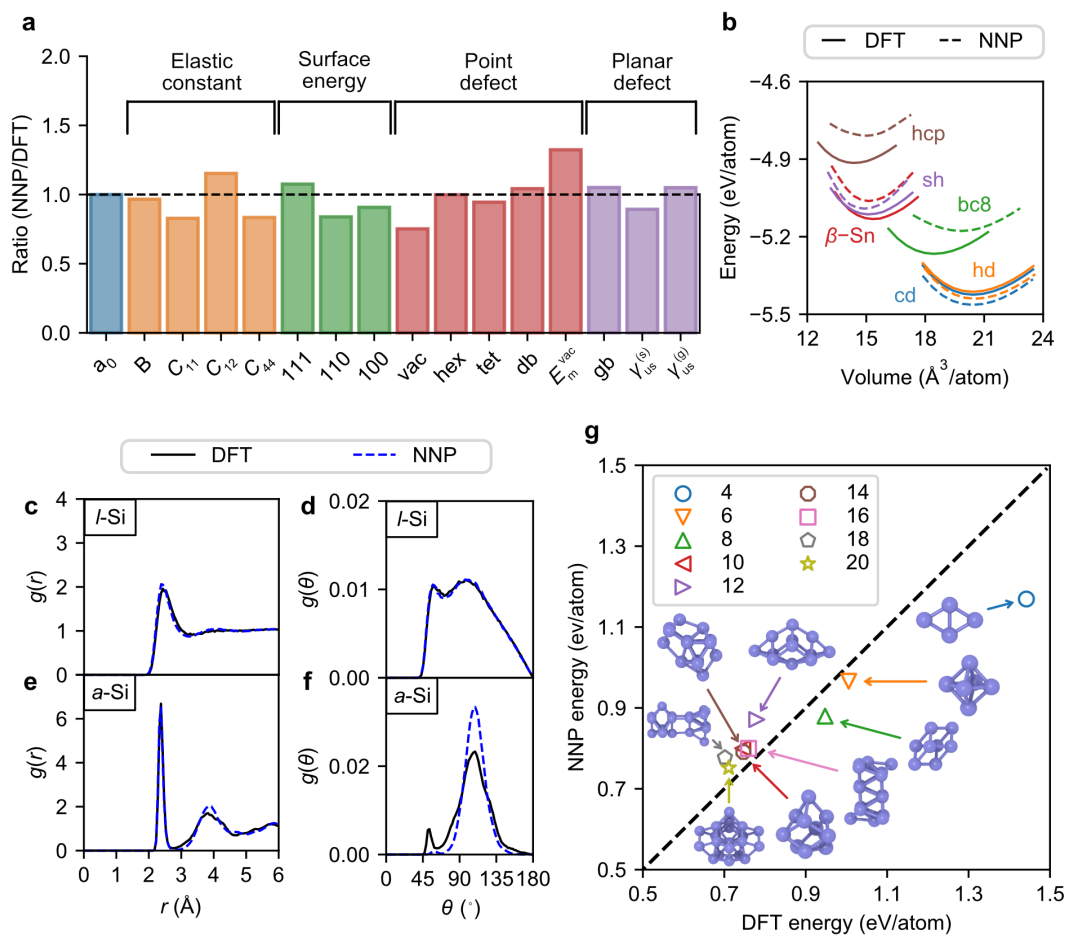

**Fig. 5** Comparison between NNP and DFT results over various properties of Si. **a** Ratios of NNP to DFT for static properties of Si in the diamond structure. Surface energies are calculated for (100)-(2×2),[47] (110)-(1×1),[48] and (111)-(3×3) reconstructions.[49] Defect formation energies for the vacancy (vac) and interstitials (hexagonal (hex), tetrahedral (tetra), and dumbbell (db)). $E_\mathrm{m}^\mathrm{vac}$ is the migration energy of the vacancy. For extended defects, gb means the (112)Σ3 grain boundary. $\gamma_\mathrm{us}^{(s)}$ and $\gamma_\mathrm{us}^{(g)}$ are unstable stacking-fault energies on shuffle and glide planes of the diamond (111) plane. **b** Equation of states for polymorphs. The abbreviations cd, hd, bc8, and sh stand for cubic diamond, hexagonal diamond, body-centered cubic, and simple hexagonal, respectively. **c** RDF and **d** angular distribution function (ADF; $g(\theta)$

of liquid Si (*l*-Si) at 2500 K. **e** RDF and **f** ADF of amorphous Si (*a*-Si) at 300 K. In **b-f**, solid and dashed lines indicate the reference DFT and NNP results. **g** Energies of even-membered nanoclusters in reference to the equilibrium diamond structure obtained by each method. Insets are structures relaxed by the NNP.

## DISCUSSION

The showcase examples on GeTe and Si in the above demonstrate that the G-metaD can produce training sets that are comparable to those that experts collected elaborately. This confirms that G-metaD can generate diverse and relevant configurations semi-automatically, which will expedite the development of MLPs by mitigating technicalities of choosing reference structures. However, a limited number of G-metaD trajectories may not provide full accuracy for every region of PES as observed in the example of Si wherein the EOS of some polymorphs is inaccurate. Therefore, we advise that practitioners augment the training set if high accuracy is necessary for specific configurations. For example, by including additional G-metaD trajectories starting with the same diamond structure but under constant pressures of 10-20 GPa, we could significantly improve EOS of high-pressure phases such as hcp and bc8 phases.

One can also utilize the G-metaD in complementing the traditional sampling style: Since MLPs are essentially interpolative algorithms, the prediction error increases rapidly with structures outside the training domain. The MD-based sampling rarely explores high-energy regions and so the trained MLP is vulnerable to failures in long-term, large-scale simulations because some atoms may visit untrained regions eventually. This can be partly resolved by a weighting scheme[50] but the present G-metaD can provide a more robust solution. That is to say, after preparing a training set based on the traditional approach, practitioners may augment the training set by adding G-metaD trajectories, which extra-samples high-energy regions relevant for the simulation. This will achieve both high accuracy and stability of subsequent simulations.

In some cases, it is useful to apply a partial-bias G-metaD in which only a few selected atoms contribute to the total bias potential. For instance, to sample various sites of an interstitial atom (self or dopant types), one can add the interstitial atom into the crystalline bulk and apply the G-metaD only to the interstitial atom. This will enhance sampling of defective structures embedded in the crystalline bulk, which would not be feasible if the biasing force drives all atoms out of the crystalline structure simultaneously. For instance, diffusion paths of Li within a solid would be sampled efficiently by the partial-bias G-metaD.[51]

About the computational cost, the G-metaD takes about three times longer than the corresponding MD in the case of 20-ps simulations of GeTe. The present implementation of G-metaD operates the client-server mode between LAMMPS and VASP, and the read-write time of wave-function files is substantial. This could be alleviated by implementing the G-metaD directly into the *ab initio* program. Another source of higher computational loads is the computation of bias forces in Eq. (2). Unlike typical metaD, the CVs in G-metaD have large dimensions of 50–100 and the bias potential in Eq. (3) is contributed by every atom in the system. As a result, the computational time of bias calculations becomes significant as the G-metaD proceeds. By reducing the dimension of the **G** vector used in G-metaD, it would be possible to increase the computational speed.

## METHODS

**Density functional theory calculations**

The reference DFT calculations are performed with Vienna *Ab initio* Simulation Package (VASP)[30] using projector augmented-wave pseudopotentials.[52] The generalized gradient approximation is used for the exchange-correlation energy of electrons.[53] In the case of GeTe, we include a parameterized van der Waals interaction.[54,55] The temperature is controlled by the Nosé-Hoover thermostat and a time step of 2 fs is used. In MD or G-metaD simulations for H:Pt(111), the cutoff energy of 350 eV and **k**-point grid of 5×5×1 are used. The G-metaD simulations with GeTe and Si are carried out with the cutoff energy of 300 and 250 eV, respectively, and the **k**-point grids are varied with a spacing of 0.4 Å$^{-1}$ to maintain the computational consistency during the large volume change.

In obtaining reference energies, forces, and stress tensors of sampled structures, we perform one-shot DFT calculations with tighter parameters such that the total energy and atomic forces converge within 1.5 meV/atom and 0.04 eV/ Å, respectively, for randomly sampled G-metaD snapshots. The resulting cutoff energy and **k**-point spacing are 400 eV and 0.3 Å$^{-1}$ for GeTe, and 350 eV and 0.157 Å$^{-1}$ for Si, respectively.

**Neural network potential**

The NNPs are trained by using SIMPLE-NN.[29] In training NNPs, the reference DFT data are split randomly into the training and validation sets with 9:1 ratio for H:Pt(111) and 19:1 ratio for GeTe and Si. We use the atom-centered symmetry function vector (**G**) to represent local environments.[23] The symmetry function vector **G** consists of radial ($G_2$) and angular components ($G_4$ and $G_5$) with cutoff radii of 3.5–8.0 Å. For training GeTe and Si, symmetry-function parameters are selected from a large pool of 233 sets by using the CUR method[56,57]. In the process of CUR, each symmetry function is penalized by rough estimates of evaluation costs based on the cutoff radius and the function type (radial or angular), thereby selecting the most cost-effective set of parameters. The CUR selection is terminated when the error ($\varepsilon$), defined in the following, drops below a certain threshold (0.001 and 0.002 for GeTe and Si, respectively):

$$\varepsilon = \|\mathbf{A} - \widetilde{\mathbf{A}}\|_F / \|\mathbf{A}\|_F \quad , \tag{5}$$

where $\mathbf{A}$ is the original feature matrix constructed from 233 parameters, and $\widetilde{\mathbf{A}}$ is the reduced feature matrix from selected parameters, and $\|\mathbf{A}\|_F$ is the Frobenius norm of matrix $\mathbf{A}$. As a result, 61, 103, and 47 symmetry functions are selected for Ge, Te, and Si, respectively. For H:Pt(111), 70 parameters with a constant cutoff of 6.0 Å are selected without applying the CUR method.

For the NNP architecture, we adopt atomic neural networks with two hidden layers. The number of nodes per hidden layer is optimized with respect to the training RMSE. As a result, each hidden layer consists of 30 nodes for H:Pt(111) and Si, and 60 nodes for GeTe. Since decorrelating the input vector benefits training quality and convergence speed, we transform the input vector by principal component analysis without dimension reduction. After the transformation, variances of the vector components are normalized by whitening.

The training is performed with momentum-based Adam optimizer[58] with minibatch (the batch size of 20), which balances between performance and computational costs. The initial learning rate is 0.0001 and reduced exponentially. The loss function ($\Gamma$) is formulated as follows:

$$\Gamma = \frac{1}{M}\sum_{i=1}^{M}\left(\frac{E_i^{\mathrm{DFT}} - E_i^{\mathrm{NNP}}}{N_i}\right)^2 + \frac{\mu_1}{3}\frac{1}{\sum_{i=1}^{M} N_i}\sum_{i=1}^{M}\sum_{j=1}^{N_i}|\mathbf{F}_{ij}^{\mathrm{DFT}} - \mathbf{F}_{ij}^{\mathrm{NNP}}|^2$$
$$+ \frac{\mu_2}{3}\frac{1}{M}\sum_{i=1}^{M}\sum_{k=1}^{3}|S_{ik}^{\mathrm{DFT}} - S_{ik}^{\mathrm{NNP}}|^2 + \frac{\mu_3}{3}\frac{1}{M}\sum_{i=1}^{M}\sum_{k=4}^{6}|S_{ik}^{\mathrm{DFT}} - S_{ik}^{\mathrm{NNP}}|^2 \quad , \tag{6}$$

where $M$ is the total number of structures in the training set, $N_i$ is the number of atoms in the $i$th structure. In Eq. (6), $E_i^{\mathrm{DFT(NNP)}}$, $\mathbf{F}_{ij}^{\mathrm{DFT(NNP)}}$, and $S_{ik}^{\mathrm{DFT(NNP)}}$ for the $i$th structure indicate the total energy, atomic force of the *j*th atom, and the *k*th component ($k$ = 1-6) of the virial stress, respectively. The scaling parameters $\mu_1$, $\mu_2$, and $\mu_3$ in Eq. (6) control the weights of the force, normal ($k$ = 1-3), and shear ($k$ = 4-6) stress terms relative to the energy term, respectively. Since the shear components are usually smaller than normal ones, employing different scaling parameters $\mu_2$ and $\mu_3$ improves accuracy in the shear modulus. The training and validation errors are similar and their RMSE values for each system are noted in the main text.

**Mahalanobis distance**

We utilize the Mahalanobis distance to classify **G** vectors in GeTe structures into a point group $\theta$ (crystal, amorphous, or liquid phase).[42] The Mahalanobis distance ($d$) measures the distance between a certain data point **x** and the center of the data point group $\theta$ in a multidimensional space.[41] It is calculated as:

$$d(\mathbf{x}, \theta) = \sqrt{(\mathbf{x} - \mu_\theta)^\mathrm{T} \Sigma_\theta^{-1} (\mathbf{x} - \mu_\theta)} \quad , \tag{7}$$

where $\mu_\theta$ and $\Sigma_\theta$ are the mean and covariance matrix of the point group $\theta$, respectively. When all axes are independent, $\Sigma_\theta$ becomes an identity matrix and the Mahalanobis distance is equal to the Euclidean one. In order to classify the unlabeled **G** vector, we first prepare reference data points for the GeTe phases from MD simulations: rock-salt crystal at 700 K, amorphous at 500 K (2 structures), and liquid at 1000 K. Then we randomly select 4,000 points from each MD simulation and label them as corresponding groups. Then, a **G** vector is classified into one of three phases ($\theta^*$) for which it has the shortest $d$:

$$\theta^* = \operatorname*{argmin}_{\theta \in \{C, A, L\}} d(\mathbf{x}, \theta) \quad , \tag{8}$$

where C, A, and L indicate crystal, amorphous, and liquid structures, respectively. In addition, if *d* of a **G** vector is longer than that of the outermost **G** vector in each phase, it remains to be unclassified. That is to say, if **x**\* is the unclassified **G** vector, the following relation holds:

$$\forall \theta, d(\mathbf{x}^*, \theta) > \max_{\mathbf{x} \in \theta}\{d(\mathbf{x}, \theta)\} \quad . \tag{9}$$

Thus, surface structures or unary phases are unclassified. We also check the consistency by using *d* as a phase classifier: *d* can map **G** vectors from crystal, amorphous and liquid into its original label with the accuracy of 92.2%, 78.8% and 100%, respectively. Most of mislabeled **G** vectors are mapped into the liquid phase due to its largest variance.

## DATA AVAILABILITY

All data generated or analyzed during this study are included in this published article (and its supplementary information files)

## CODE AVAILABILITY

The code for G-metaD is available upon request.

## ACKNOWLEDGMENTS

This work was supported by Samsung Electronics Co., Ltd. The computations were carried out at Korea Institute of Science and Technology Information (KISTI) National Supercomputing Center (KSC-2020-CRE-0125)

## AUTHOR CONTRIBUTIONS

D.Y. put forward the original idea of G-metaD and further refined the method with J.J., W.J. and S.H. S.H. organized the whole project. All the authors participated in writing the manuscript.

## COMPETING INTERESTS

The authors declare no competing interests.